\documentclass[9pt,twocolumn,twoside]{optica}
\setboolean{shortarticle}{false}
\setboolean{minireview}{false}

\usepackage{xr}

\newcommand{\ua}{\uparrow}
\newcommand{\da}{\downarrow}

\newcommand{\bs}{\boldsymbol}

\title{Theory of Kerr Instability Amplification}

\author[1,*]{M. Nesrallah}
\author[1]{G. Vampa}
\author[1]{G. Bart}
\author[1,2]{P. B. Corkum}
\author[1]{C. R. McDonald}
\author[1]{T. Brabec}
\affil[1]{Department of Physics, University of Ottawa, Ottawa, ON K1N 6N5 Canada}
\affil[2]{National Research Council of Canada, Ottawa, Ontario, Canada K1A 0R6}

\affil[*]{Corresponding author: mnesr024@uottawa.ca}

\ociscodes{}

\begin{abstract}
A new amplification method based on the optical Kerr instability is suggested and theoretically analyzed, with emphasis on the near- to mid-infrared wavelength regime. Our analysis for CaF$_2$ and 
KBr crystals shows that one to two cycle pulse amplification by 3-4 orders of magnitude in the wavelength range from $1-14\,\mu$m is feasible with currently available laser sources. At $14\,\mu$m 
final output energies in the 50 $\mu$J range are achievable corresponding to about 0.2-0.25\% of the pump energy. The Kerr instability presents a promising process for the amplification of ultrashort 
mid-infrared pulses. 
\end{abstract} 

\setboolean{displaycopyright}{true}

\begin{document}

\maketitle

\section{Introduction}

\noindent
Research in strong field physics and attosecond science has triggered the need for high intensity ultrashort laser sources in the mid-infrared \cite{ghimire2011, schiffrin2013, gattass2008}. Currently, 
the most common generation and amplification methods are based on the second order nonlinearity, such as optical parametric amplifiers (OPAs) \cite{malevich2013,schmidt2014,manzoni2016}; recently the 
potential for single cycle infrared pulse generation by difference frequency generation has been demonstrated \cite{krogen2017}. 

Although OPAs are currently the leading technology for ultrashort mid-infrared pulse amplification, their development is challenging. For their efficient operation a series of stringent conditions must be met 
which are intimately connected to the properties of second order nonlinear crystals. Amplification of single-cycle pulses either requires thin crystals (reducing the efficiency), or low dispersion across 
a spectrum covering the frequencies of the three interacting waves. Moreover, many second-order nonlinear crystals absorb light in the mid-infrared, and moderate damage thresholds also present a limitation. 

\begin{figure}
\hspace*{-0.8cm}
\includegraphics[scale=0.395]{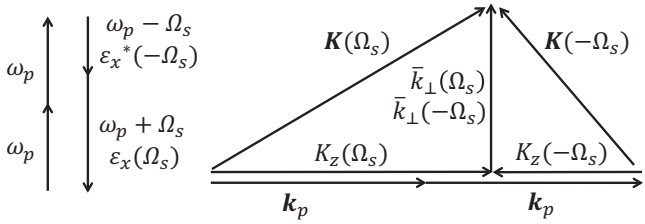}
\caption{Schematic of Kerr instability amplification (KIA). Left: parametric four wave mixing process of the type $ 2 \omega_p - (\omega_p \pm \Omega_s) = \omega_p \mp \Omega_s$ where $\omega_p$ is the 
pump frequency and $-\omega_p < \Omega_s < \omega_p$ is the seed frequency. Right: there exist transverse wavevectors for which unstable behavior occurs; gain is maximum for the transverse wavevector 
$\bar{k}_{\perp}(\Omega_s)$. The instability evolves as $\varepsilon_x(\Omega_s) = \exp(i(\omega_p + \Omega_s) t -i\mathbf{K}(\Omega_s) \mathbf{x})$ and $\varepsilon_x^*(-\Omega_s) = \exp(-i(\omega_p - 
\Omega_s) t + i\mathbf{K}(-\Omega_s) \mathbf{x})$. Here, $\mathbf{K}(\Omega_s) = (\bar{k}_{\perp}(\Omega_s), 0, K_z(\Omega_s))$. Phase matching of the instability is automatically fulfilled as the wavevector 
of the instability $\mathbf{K}(\Omega_s)$ fulfills the relation $\mathbf{K}(\Omega_s) = 2 \mathbf{k}_p + \mathbf{K}(-\Omega_s)$. For a more detailed discussion, see Eq. (\ref{pw}) and below. 
\label{fig0} }
\end{figure}

Here we introduce an alternative concept for mid-infrared amplification based on the Kerr nonlinearity which we call Kerr instability amplification (KIA). In a Kerr nonlinear material parametric four 
wave mixing processes of the type $\omega_p + \omega_p - (\omega_p \pm \Omega_s) = \omega_p \mp \Omega_s$ occur, during which two photons $\omega_p$ of the pump field are converted into fields 
$\varepsilon_x(\Omega_s)$ and $\varepsilon_x^*(-\Omega_s)$ with photon energies shifted to the red and blue side of $\omega_p$ by $\Omega_s$, see Fig. \ref{fig0}. We find that for a wide range of seed frequencies 
in the interval $-\omega_p < \Omega_s < \omega_p$ there exist transverse wavevectors $k_{\! \perp}$ for which unstable behavior occurs that results in exponential growth. As the transverse wavevector for 
maximum amplification $\bar{k}_{\! \perp}(\Omega_s)$ is finite, emission is noncollinear with the pump pulse. Mathematically, the instability emerges from a coupling between the wave equations for 
$\varepsilon_x(\Omega_s)$ and $\varepsilon^*_x(-\Omega_s)$. This yields a second order Mathieu-type equation that contains unstable solutions. The wavevectors of the instability $\mathbf{K}(\pm \Omega_s)$ 
fulfill the relation $\mathbf{K}(\Omega_s) = 2 \mathbf{k}_p + \mathbf{K}(-\Omega_s)$, see Fig. \ref{fig0}, so that phase matching is automatically fulfilled. Outside the unstable range, the phases of 
$\varepsilon_x(\Omega_s)$ and $\varepsilon_x^*(-\Omega_s)$ are mismatched and regular four wave mixing dynamics ensues. Through this instability a Kerr nonlinear material irradiated by a high intensity 
pump pulse can act as an amplifier for a noncollinear seed pulse. 

It is well known that intense laser pulses propagating in Kerr nonlinear materials result in self focusing, breakup and the formation of stable filaments. From these filaments conical emission occurs -- 
the emission of broad band radiation at a frequency dependent angle to the filament; for a review see Ref. \cite{couairon2007}. KIA is similar to conical emission, however it occurs long before filamentation 
happens, in the limit where the Kerr nonlinearity has not substantially modified the pump pulse. 

A complete characterization of KIA requires knowledge of the complex wavevector of the instability in the whole spectral and transverse wavevector ($k_{\! \perp}$) domain. This is obtained here by an extended 
linear stability analysis. In the limits of $\Omega_s = 0$ and $k_{\! \perp} = 0$ KIA gain reduces to the well known cases of filamentation instability \cite{bespalov1966} and modulation instability 
\cite{agrawal2012}, respectively. 

Our theoretical results are used for a proof-of-principle feasibility analysis of KIA on the basis of two infrared materials, CaF$_2$ and KBr. We find that amplification of seed wavelengths in the range 
from $1\,\mu$m to 14$\,\mu$m is possible with amplification factors and seed pulse energies that are competitive with OPAs; over most of the wavelength range single-cycle pulse amplification is supported. 
We believe that with optimization and further progress in infrared pump laser source development, KIA has the potential to become a versatile tool for ultrashort pulse amplification in the infrared.

\section{Theory of Kerr instability amplification}

\noindent
A summary of all the parameters and definitions used in our derivation is given in the supplement \cite{supplement}.  
Our analysis of KIA starts from Maxwell's equations for a Kerr ($\chi^{(3)}$) nonlinear material, cast into a vector wave equation for the electric field, $\mathbf{E}(\mathbf{x},t) = E_p \hat{\mathbf{x}} 
\exp(i\omega_p t-i k_{p} z) + \boldsymbol{\varepsilon}(\mathbf{x},t) + {\rm c.c.}$; the electric field is chosen as a superposition of a pump continuous wave (cw) polarized along $x$ and a 
perturbation, $\boldsymbol{\varepsilon}$. Here $E_p$ is the pump electric field strength, $\omega_p$ is the pump laser frequency, and $k_{p}$ is the pump wavevector defined below. 

Inserting the ansatz into the vector wave equation and keeping only terms $\mathrm{O}(\boldsymbol{\varepsilon})$ gives 
\begin{align} 
& \left( \partial_z^2 + \boldsymbol{\nabla}^2_{\perp} - \boldsymbol{\nabla} (\boldsymbol{\nabla\cdot}) - {\partial_t^2 \over c^2}   n^2 \! \star \right) \boldsymbol{\varepsilon} 
= \frac{E_p^2\partial_t^2}{c^2} \chi^{(3)} \! \star \mathbf{P}(\boldsymbol{\varepsilon}) \text{} \label{vwaveqt}
\end{align}
with $ \mathbf{P} = 2\left(\boldsymbol{\varepsilon}+ 2\hat{x} \varepsilon_x\right)  + \left(\boldsymbol{\varepsilon}^* + 2\hat{x} \varepsilon_x^*\right) \exp[2i ( \omega_pt-k_{p} z ) ]$; the star symbol, $\star$, represents convolution. The cw field is a solution of the 
vector wave equation for $k_{p} = (n_p^2+n_{n}(\omega_p))^{1/2} \omega_p/c$ and drops out of Eq. (\ref{vwaveqt}). Here, $n(\omega)$ is the linear refractive index defined in the frequency domain, $n_p = n(\omega_p)$, and $n_{n} = 3\chi^{(3)}E_p^2=
n_2I_p$ is the nonlinear refractive index with $n_2$ the optical Kerr nonlinearity coefficient, and $I_p$ the pump intensity. We assume $n_n$ constant, which is a reasonable approximation for frequencies much smaller than the material bandgap. 
Next, we define $\boldsymbol{\varepsilon} = \mathbf{v}(\mathbf{x},t) \exp (i \omega_p t -i k_{p} z)$ and perform a Fourier transform of Eq. (\ref{vwaveqt}) from coordinates $x,y,t$ to $\mathbf{k}_{\perp}, 
\Omega = \omega-\omega_p$, where $\mathbf{k}_{\perp} = (k_x,k_y)$ defines the transverse wavevector. The Fourier transform is denoted as $\hat{F}(\varepsilon_x) = \tilde{\varepsilon}_x(z,\omega, \mathbf{k}_{\perp}) 
= \tilde{v}_x(z,\Omega,\mathbf{k}_{\perp}) \exp (-i k_{p} z)$. The Fourier transformed wave equation is 
\begin{align} 
\left[ (\partial_z - ik_{p})^2 + k_{v}^{2} - k_{\perp}^2 \right] \tilde{v}_x = - k_{n}^2 \, \tilde{v}_{x(-)}^* , 
\label{vwaveq}
\end{align}
where $k_{v}^{2}(\omega) = k^2(\omega) + 2 k^2_{n}(\omega)$ is the wavevector experienced by the perturbation; it is composed of a linear contribution, $k = n\omega/c$, and a nonlinear wavevector, $k_{n} = 
n_{n}^{1/2} \omega/c$; $k_{\perp}^2=k_x^2+k_y^2$ is the transverse wavevector squared. Also, we use the notation $\tilde{v}_{x}^*(-\Omega) = \tilde{v}_{x(-)}^*$. Note that the wavevector $k_{v}$ of the 
perturbation contains twice the nonlinearity of the pump wavevector, $k_{p}$, which comes from the first two terms in $\mathbf{P}$ defined below Eq. (\ref{vwaveqt}). Also, we have assumed small nonlinearity, 
$n_{n} / n^2 \ll 1$, for which the $\boldsymbol{\nabla} (\boldsymbol{\nabla\cdot})$ operator coupling different polarization directions in Eq. (\ref{vwaveqt}) can be neglected. Materials for 
which $n \rightarrow 0$ violate this assumption and require separate consideration \cite{alam2016}.

The equation for $\tilde{v}^*_{x(-)}$ is obtained by taking the complex conjugate of Eq. (\ref{vwaveq}) and by replacing $\Omega \rightarrow -\Omega$ in all $\Omega$-dependent functions, 
\begin{align} 
\left[ \left( \partial_z + ik_{p} \right)^2 + k_{ v(-)}^{2} - k_{\perp}^2 \right] \tilde{v}^*_{x(-)} = - k^2_{n (-)} \tilde{v}_x \text{.}
\label{vwaveq-} 
\end{align}
Here, $k^{2}_{v (-)} = k_{v}^2(\omega_p-\Omega)$, and the minus in $k^2_{n (-)}$ has the same meaning. 

In order to make further progress the sign swapped functions need to be specified. To this end, they need to be split in even/odd parts that are symmetric/antisymmetric with regard to sign change. We start 
with $k_{v}(\omega)$ and introduce $\eta = \sqrt{n^2 + 2n_n} \approx n+n_n/n$ and $\eta_p = \eta(\omega_p)$. The refractive index can be recast into $ \eta(\omega) = \eta_p + \Delta \eta(\Omega) $, where $\Delta 
\eta(\Omega) = \eta_{g}(\Omega) + \eta_{u}(\Omega)$ is split into even and odd parts, $\eta_{g,u} = {1\over 2} [ \Delta \eta(\Omega) \pm \Delta \eta(-\Omega) ]$, so that $\eta_{g}(-\Omega) = \eta_{g}(\Omega)$ 
and $\eta_{u}(-\Omega) = - \eta_{u}(\Omega)$. Inserting these definitions we obtain $k_{v} = k_{v}(\omega_p) + D_{g} + D_{u}$, where 
\begin{subequations}
\label{dispersion}
\begin{align}
D_{g}(\Omega) & = \frac{\eta_{g}(\Omega) \omega_p + \eta_{u}(\Omega) \Omega} {c} \text{,} \label{Dg} \\
D_{u}(\Omega) & = \frac{(\eta_p + \eta_{g}(\Omega) ) \Omega + \eta_{u}(\Omega) \omega_p} {c} \label{Du} \text{.} 
\end{align}
\end{subequations}
Using the above definitions, the sign flipped wavevector is given by $k_{v (-)} = k_{v}(\omega_p) + D_{g} - D_{u}$. In the absence of nonlinearity, $\eta_{p,g,u} \rightarrow n_{p,g,u}$, $k_v \rightarrow k$, 
and $D_{g,u}$ become the linear even and odd dispersion terms. For $\Omega/\omega_p \ll 1$, $n_{u} \approx n_p' \Omega$ and $n_{g} \approx n_p'' \Omega^2/2$ so that to lowest order we obtain from 
Eq. (\ref{dispersion}) $D_{g} \approx (\beta_2/2) \Omega^2$ and $D_{u} \approx \beta_1 \Omega$ with $\beta_1 = [dk/d\omega](\omega_p) = (n_p+n_p'\omega_p)/c$ group velocity and $\beta_2 = [d^2k/d\omega^2](\omega_p) 
= (n_p''\omega_p + 2n_p')/c$ group velocity dispersion; prime and double prime denote first and second frequency derivative, respectively. 
Since we treat $n_n$ as constant, the sign flip operation for 
the nonlinear wavevector is trivial, $k^2_{n(-)} = n_{n} (\omega_p - \Omega)^2 /c^2$. 

Using Eq. (\ref{vwaveq-}) to eliminate $\tilde{v}^*_{x(-)}$ in Eq. (\ref{vwaveq}) results in a fourth order differential equation. 
Inserting the Ansatz $\tilde{v}_x \propto \exp(iK_vz)$ with $K_v(\Omega)$ a complex wavevector yields the quartic equation 
\begin{align} 
~& \left[ (K_v^2 - D_{\rm u}^{2} \sigma^2) + (D_u^2 - k_{\rm p}^{2}) (\sigma^2-1) + k_{\perp}^2 \right]^2 -  \nonumber \\
- & 4k_{\rm p}^2\left(K_v + D_{\rm u} \sigma \right)^2 - k^2_{\rm n} k^2_{\rm n(-)}=0   \text{}
\label{quartic} 
\end{align}
with $\sigma(\Omega) = (k_{v}(\omega_p)+D_{g}) / k_{p}$. By using $k_{v}(\omega_{p}) \approx k_{p} + k^2_{n}(\omega_{p})/(2k_{p})$ we obtain the approximate expression $\sigma^2 - 1 \approx 
(k_n(\omega_p) / k_p)^2 + 2 D_{\rm g} / k_{\rm p}$ for later use. 
The dominant part of the solution is given by the the second term, which gives $K_v \approx -D_{\rm u} \sigma$. As a result, we can approximate in the first term of 
Eq. (\ref{quartic}) $K_v^2 - (D_{u} \sigma)^2 \approx -2D_u \sigma (K_v + D_u \sigma)$. This amounts to neglecting backward propagating solutions and results in a reduction to a quadratic equation, 
\begin{align}
& 4 (K_v + \sigma D_{u})^2 \left( k_{p}^2  - ( \sigma D_{u} )^2 \right) - 4 (K_v + \sigma D_{u}) \times \nonumber \\
& \times \sigma D_{u} \left( \kappa_{\! \perp}^2 - k^2_{\!\perp } \right) - \left( \kappa_{\! \perp}^2 - k^2_{\!\perp} \right)^2 + \left( k_{n} k_{n(-)} \right)^2 = 0 \text{.}
\label{quadratic}
\end{align}
Here, $\kappa_{\! \perp}^2(\Omega) = (k_{p}^2 -D_{u}^2) (\sigma^2 - 1)$. Solution of Eq. (\ref{quadratic}) yields $K_v = K_{u}(\Omega) + K_{g}(\Omega)$ with 
\begin{subequations}
\label{KOm}
\begin{align}
K_{u}(\mathbf{k}_{\perp},\Omega) & = - \sigma D_{u} \left[ 1 - {1\over 2}{ \kappa_{\! \perp}^2 - k^2_{\! \perp} \over k_{p}^2  - ( \sigma D_{u} )^2 } \right]  \label{KOmu} \\
K_{g}(\mathbf{k}_{\perp},\Omega) & = - {1\over 2} { k_{p} \sqrt{ \left( \kappa_{\! \perp}^2 - k^2_{\! \perp} \right)^2 - \delta_{\! \perp}^4 } \over k_{p}^2  - ( \sigma D_{u}  )^2  } \label{KOmg} \text{;}
\end{align}
\end{subequations}
$\delta_{\! \perp}^2$ is defined below. In the appropriate limits \cite{supplement}, Eq. (\ref{KOmg}) goes over into the temporal modulation instability \cite{agrawal2012}, and the spatial filamentation instability 
\cite{bespalov1966}.
Note that the quadratic equation corresponds to a second order Mathieu-type differential equation that supports unstable solutions. When the argument of the square root in $K_{g}$ is negative, exponential growth 
happens with intensity gain $g = - 2 \mathrm{Im}(K_{g})$. In the limit of $k_{p}^2  = ( \sigma D_{u}  )^2$, which occurs for $\Omega \approx \pm \omega_p $, the quadratic equation (\ref{quadratic}) reduces to a 
linear equation and $K$ becomes real; this has to be treated separately. For each frequency $\Omega$ the gain $g$ is maximum at transverse wavevector 
\begin{align}
\bar{k}_{\! \perp}(\Omega) = 
\begin{cases} 
\kappa_{\! \perp} & \mathrm{for} \,\, \kappa_{\! \perp}^2 \ge 0 \\
0  & \mathrm{for} \,\, \kappa_{\! \perp}^2 < 0
\end{cases}
\label{kpmax}
\end{align}
and is denoted by $\bar{g} = g(k_{\! \perp} = \bar{k}_{\! \perp}(\Omega), \Omega )$ with
\begin{align}
\bar{g}(\Omega) = 
\begin{cases} 
{ k_{p} \sqrt{ \delta_{\! \perp}^4 - \left( \kappa_{\! \perp}^2 - \bar{k}^2_{\! \perp} \right)^2 } \over k_{p}^2  - ( \sigma D_{u}  )^2 } & \mathrm{elsewhere} \\
0 & \mathrm{for} \,\, \kappa_{\! \perp}^2 < 0, \kappa_{\! \perp}^4 > \delta_{\! \perp}^4 ; 
\end{cases}
\label{gmax}
\end{align}
As $g( k_{\! \perp}^2 = \bar{k}_{\! \perp}^2 \pm \delta_{\! \perp}^2 ) = 0$, the transverse wavevector halfwidth squared over which KIA gain occurs for a given $\Omega$ is given by 
\begin{align}
\delta_{\! \perp}^2(\Omega) = { k_{n} k_{n(-)} \over k_{p} } \sqrt{ k_{p}^2  - (\sigma D_{u} )^2 } \text{}
\label{kwidth}
\end{align}
The relation $k_{\! \perp}^2 = \bar{k}_{\! \perp}^2 \pm \delta_{\! \perp}^2$ defines curves in the $k_{\! \perp}-\Omega$ plane at which gain disappears. The curve defined by the expression with the minus sign 
exists only for $\kappa_{\! \perp}^2 \ge \delta_{\! \perp}^2$. 

\section{Kerr instability amplification in the plane wave limit}

\subsection{Theory}

\noindent
A seed plane wave 
\begin{align}
\tilde{v}_x(z=0) = (2\pi)^{3/2}E_{\rm s} \delta(k_x - \bar{k}_{\! \perp s} ) \delta(k_y) \delta(\Omega-\Omega_s)\label{pwfourier}
\end{align}
experiences maximum gain according to the above relations; here, $\bar{k}_{\! \perp s} = \bar{k}_{\! \perp}(\Omega_s)$. After material length $l$ the electric field is determined by the inverse Fourier 
transform of $\tilde{v}_x(0) \exp(iK_vl)$ which yields 
\begin{align}
\varepsilon_x(\mathbf{x},t) & = E_{s} \exp \left( {1\over 2} \bar{g}(\Omega_s) l - i \mathbf{K}_s \mathbf{x} + i \omega_s t \right)  \text{.}
\label{pw}
\end{align}
Here, $\mathbf{K}_s = \mathbf{K}(\Omega_s) = ( \bar{k}_{\! \perp s}, 0, K_{zs} ) $ is the seed wavevector, $K_{zs} = K_{z}(\Omega_s) = k_{p} + [\sigma D_{u}](\Omega_s)$, $\mathbf{x} = (x, y,z=l)$, 
$\omega_s = \omega_p + \Omega_s$, and $E_s$ is the seed electric field strength. We find that optimum amplification takes place when the seed propagation axis lies on a cone around the pump wavevector with 
half-angle 
\begin{align}
\theta_s = \theta(\Omega_s) = \arctan \left( \bar{k}_{\! \perp s} / K_{zs}  \right)
\label{thetas}
\end{align}
Note that $\theta_s$ is related to but not the same as the conical emission angle. Conical emission grows out of noise and operates in the regime of filamentation where the pump pulse has been drastically 
modified through Kerr nonlinearity and other processes. Seeded amplification happens over distances long before filamentation sets in. 

Further, we would like to point out that KIA is automatically phase matched, unlike conventional three- or four-wave mixing processes, see also the schematic in Fig. \ref{fig0}. The space 
dependent phase of the perturbation terms is $\tilde{v}_{x} \propto \exp(-i \mathbf{K}(\Omega_s) \mathbf{x} )$ and $\tilde{v}_{x(-)}^* \propto \exp(i \mathbf{K}(-\Omega_s) \mathbf{x} )$. As $ 
\bar{k}_{\! \perp}(\Omega_s) = \bar{k}_{\! \perp}(-\Omega_s)$ and $K_z(- \Omega_s) = - K_z(\Omega_s)$, the left and right hand side of Eq. (\ref{vwaveq}) are automatically phase matched. This is not the 
case outside the instability regime, where $K_g$ becomes real, as $K_g(\Omega_s) = K_g(-\Omega_s)$, which is the conventional regime of four wave mixing. 

\subsection{Discussion of results}

\begin{figure}[t]
\includegraphics[scale=0.48]{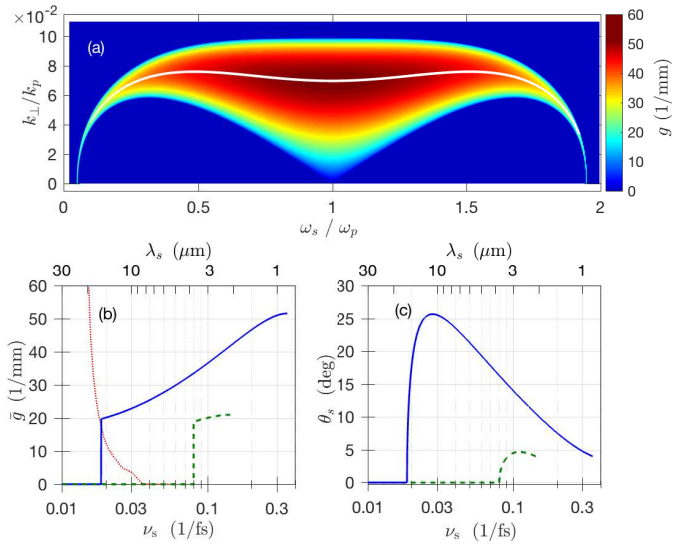}
\caption{Plane wave amplification in CaF$_2$ crystal with $n_2 = 2 \times 10^{-16}$ cm$^2$/W; $I_p = 50$ TW/cm$^2$. (a) g versus $\omega_s/\omega_p$ and $k_{\! \perp}/k_{p}$; $\lambda_p=0.85\,\mu$m; 
white line indicates $\bar{k}_{\! \perp}$. (b) Maximum gain $\bar{g} = g(\bar{k}_{\! \perp})$ versus $\nu_s$ (bottom) and $\lambda_s$ (top); red dotted line represents absorption. (b)-(c) $\lambda_{p} = 0.85, 2.1 
\,\mu$m corresponds to blue full, green dashed curves, respectively. (c) $\theta_s$ from Eq. (\ref{thetas}) versus $\nu_s$ and $\lambda_s$. 
\label{fig1}}
\end{figure}

\begin{figure}[t]
\includegraphics[scale=0.48]{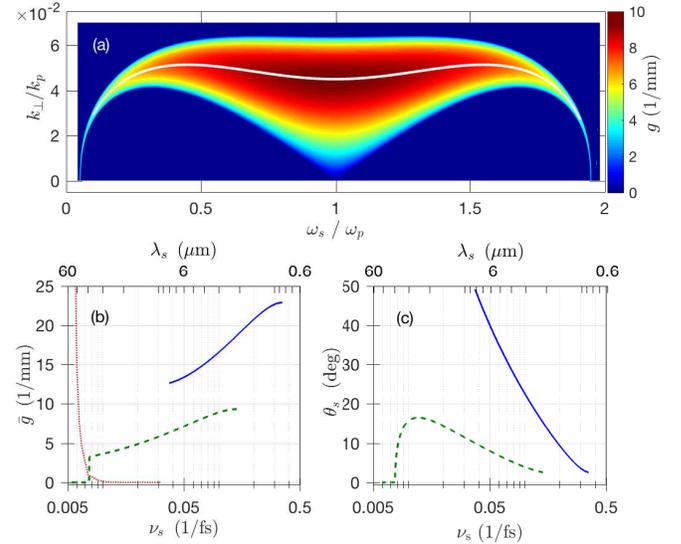}
\caption{ Plane wave amplification in KBr crystal with $n_2 = 6 \times 10^{-16}$ cm$^2$/W; $I_p = 8$ TW/cm$^2$. (a) g versus $\omega_s/\omega_p$ and $k_{\! \perp}/k_{p}$; $\lambda_p=2.1\,\mu$m; white 
line indicates $\bar{k}_{\! \perp}$. (b) Maximum gain $\bar{g}$ versus $\nu_s$ (bottom) and $\lambda_s$ (top); red dotted line represents absorption. (b)-(c) $\lambda_{p} = 0.85, 2.1 \,\mu$m corresponds to blue 
full, green dashed lines, respectively. (c) $\theta_s$ from Eq. (\ref{thetas}) versus $\nu_s$ and $\lambda_s$. \label{fig2} }
\end{figure}

\noindent
Equations (\ref{KOm}) - (\ref{thetas}) characterize KIA over the whole frequency and transverse wavevector space. In the following, these equations are discussed on the basis of CaF$_2$ and KBr 
in Figures \ref{fig1} and \ref{fig2}, respectively; we chose two different pump wavelengths $\lambda_p = 0.85, 2.1\,\mu$m. The CaF$_2$ crystal has a transmission window from $0.3-8\,\mu$m \cite{handbook1991}, 
$n_2 = 2 \times 10^{-16}$ cm$^2$/W \cite{milam1997}, and $n$ is taken from Ref. \cite{malitson1963}. The KBr crystal transmits from $0.25-25\,\mu$m \cite{handbook1991}, $n_2 = 6 \times 10^{-16}$ cm$^2$/W 
\cite{desalvo1996}, and $n$ is taken from Ref. \cite{li1976}. The Kerr nonlinear index has a maximum for frequencies around the bandgap, and decreases on the infrared side asymptotically towards the zero-frequency 
limit \cite{bahae1991}. In the wavelength range of interest here, far away from the bandgap, $n_2$ undergoes little variation. 

In Fig. \ref{fig1}(a) the intensity gain profile $g$ from Eq. (\ref{KOmg}) is plotted versus $\omega_s/\omega_p$ and $k_{\! \perp}/k_{p}$; the full white line represents $\bar{k}_{\! \perp}$; pump wavelength 
$\lambda_p = 2 \pi c / \omega_p = 0.85\,\mu$m and pump intensity is $I_p=50$ TW/cm$^2$. The validity of the analytical results has been tested by comparison to a numerical solution of wave equation (\ref{vwaveq}); 
they are found to be in excellent agreement. Amplification occurs over a wide spectral range from $0.45-15\,\mu$m. Gain terminates along two curves which are defined by the relation discussed below 
Eq. (\ref{kwidth}). 

In Fig. \ref{fig1}(b) the maximum gain $\bar{g}$ is shown on the infrared side versus seed frequency $\nu_s$ (bottom axis) and seed wavelength $\lambda_s$ (top axis); the two pump wavelengths $\lambda_{p} = 0.85, 
2.1 \,\mu$m correspond to the blue full and green dashed curves, respectively in \ref{fig1}(b) and (c). Maximum gain reaches a global maximum when pump and seed frequency are equal and drops towards longer 
wavelengths. Further, $\bar{g}$ increases with pump frequency. For $\lambda_{p} = 0.85 \,\mu$m the gain is still substantial at $\lambda_s = 15\,\mu$m; amplification ($\exp(\bar{g}l)$) by more than 4 orders of 
magnitude can be obtained in a $l=0.5$ mm long crystal. Note that gain and absorption balance each other at $\lambda_s=20\,\mu$m. As a result, the medium becomes transparent in the presence of the pump beam. 
For $\lambda_{p} = 2.1\,\mu$m the gain extends only over a narrow spectral interval. The reason for this behavior becomes clear from Fig. \ref{fig1}(c), where the angle for maximum amplification, Eq. (\ref{thetas}), 
is plotted for the same two pump wavelengths. 

For $\lambda_{p} = 2.1 \,\mu$m, $\theta_s$ reaches a maximum close to the pump wavelength and then drops to zero. This property arises from the functional form of $n(\omega)$. The angle $\theta_s$ depends on 
$\bar{k}_{\! \perp}$ which depends on $\kappa_{\! \perp}^2 \propto \sigma^2-1 \approx n_n/n_p^2 + (2/n_p)(\eta_g + \eta_u \Omega_s / \omega_p)$. Depending on the material and $\lambda_p$, the two terms $\eta_g$ 
and $\eta_u \Omega_s / \omega_p$ can have opposite or equal signs. In this particular case, they are of opposite sign and comparable magnitude, so that for decreasing $\nu_s$, $\kappa_{\perp}^2$ becomes 
negative. From Eqs. (\ref{kpmax}) and (\ref{gmax}) we see that then $\bar{k}_{\! \perp} = \bar{g} = 0$ so that both gain and $\theta_s$ become zero. A similar behavior can be seen for $\lambda_p = 0.85\,\mu$m, 
however stretched out over a wider spectral interval.  

In Fig. \ref{fig2}, the results for a KBr crystal are shown for a pump intensity $I_p=8$ TW/cm$^2$; same line styles as in Fig. \ref{fig1} are used. The intensity gain profile $g$ in Fig. \ref{fig2}(a) is plotted 
for $\lambda_p=2.1\,\mu$m. In contrast to CaF$_2$, in KBr, $\lambda_p=2.1\,\mu$m works very well and gain extends up to twice the transmission window, see also Fig. \ref{fig2}(b). The maximum gain is still 
substantial at the edge of the transmission window, see the red dotted line. There amplification of more than four orders of magnitude can be achieved over a crystal length $l=2$ mm. 
For $\lambda_p=0.85\,\mu$m, the gain is confined to a narrower spectral range (up to $8\,\mu$m). The reason becomes clear from Fig. \ref{fig2}(c). 

For $\lambda_p=0.85\,\mu$m, the angle rises sharply for increasing seed wavelength. This comes from the fact that both terms in $\sigma^2-1$ carry the same sign. Here, the gain terminates when the denominator in Eq. 
(\ref{KOmg}) goes to zero for $k_{p} = \sigma D_{u}$. By contrast for $\lambda_p=2.1\,\mu$m, the signs are again different and we see a similar behavior as in Fig. \ref{fig1}(c). Clearly, $n(\omega)$ strongly influences 
KIA and therefore presents a critical design parameter. 


\section{Kerr instability amplification of finite pulses}

\noindent
In extension of our plane wave analysis above, we explore KIA of finite pulses in a noncollinear setup with seed and pump pulses inclined at the optimum gain angle $\theta_{\rm s}$.
We also expect KIA of Bessel-Gaussian seed pulses to work well, as the KIA profile is of Bessel-Gaussian nature. This will be subject to future research. 

\subsection{Theory}

\noindent
Our analysis relies on assuming a pump plane wave. This is justified, as long as the pump pulse is wider than the seed pulse so that its intensity varies weakly over the seed pulse. The seed pulse is assumed 
to be inclined at $\theta_{\rm s}$ along $x$ with a Gaussian spatial and temporal profile and field strength $E_s$; the spatial and temporal $1/e^2$-widths are $\mathrm{w}_x(0) = \mathrm{w}_x = 2 / \Delta_{x}$, 
$\mathrm{w}_y(0) = \mathrm{w}_y = 2 / \Delta_{y}$ and $\tau = \tau(0)$, respectively. The initial Gaussian seed pulse in the Fourier domain is given by 
\begin{align}
\tilde{v}_x(0) = {2^{3/2} E_s  f(\Omega)\over \Delta_x \Delta_y \Delta_\omega} \exp \left( - \left( \frac{k_x - \bar{k}_{\perp s}}{\Delta_x} \right)^2  - \left(\frac{ k_y}{\Delta_y}\right)^2 \right)  \text{,}
\label{vseed0}
\end{align}
where $f = \exp ( - (\Omega - \Omega_s)^2 / \Delta_{\omega}^2)$ with $\Delta_{\omega}(0) = \Delta_{\omega} = 2 / \tau$. As the transverse wavevector of maximum amplification $\bar{k}_{\! \perp}(\Omega)$ varies as a 
function of frequency, (transverse) beam center and amplification maximum move increasingly apart with growing $|\Omega - \Omega_s|$. In the strong amplification limit the transverse beam center will align with the 
amplification maximum, resulting in an angular chirp \cite{gu2004}, i.e. different frequency components have slightly different transverse wavevector centers. The amplified pulse spectrum can be approximately 
evaluated analytically by Taylor expanding the gain $g$ about $\bar{k}_{\! \perp}(\Omega)$; to leading order this results in a Gaussian intensity amplification profile, where
\begin{align}
g \approx \bar{g} - g_2 \, (k_x - \bar{k}_{\! \perp})^2 \text{,} \,\,\,\,\,\, g_2  = { 2k_{p} \bar{k}_{\! \perp}^2 \over \delta_{\! \perp}^2 (k_{p}^2  - (\sigma D_{u} )^2) } \text{.}
\label{intamp}
\end{align}
The gain only modifies the $k_x$ pulse profile. Together with Eq. (\ref{KOmu}) we obtain the Fourier beam amplitude after amplifier length $l$ 
\begin{align}
~ & \tilde{v}_x(\mathbf{k}_{\perp},l,\Omega) = \tilde{v}_x(0) \exp\left(-i \sigma D_u l + \frac{1}{2} \bar{g} l\right) \exp\left(-\frac{i}{2} \alpha l k_y^2 \right) \nonumber \\
~ & \times \exp \left( -\frac{l}{2} (g_2 + i \alpha) (k_x- \bar{k}_{\perp})^2 - i \alpha l \bar{k}_{\perp} (k_x- \bar{k}_{\perp}) \right)
\text{,}
\label{epslft}
\end{align}
where $\alpha(\Omega) = \sigma D_{u} / (k_{p}^2  - ( \sigma D_{u} )^2)$. Propagation in free space after the amplifier for a length $l_1$ is not considered here; it can be accounted for by multiplying Eq. (\ref{epslft}) 
with the factor $\exp\{-i l_1 [ k(\Omega) - (1/2) k_{\!\perp}^2 / k(\Omega) ] \}$. 

Inverse Fourier transform with regard to $\mathbf{k}_{\perp}$ gives a complex shifted Gaussian beam 
\begin{align}
\tilde{v}_x(x,y,l,\Omega) & = \frac{ E_s \tau \mathrm{w}_x \mathrm{w}_y } {\sqrt{2 q_x q_y }} f(\Omega)  \exp \left( ({\gamma\over2} -i\varkappa) l + i \bar{k}_{\! \perp s} x \right) \nonumber \\
~& \times \exp \left( - { (x-x_c)^2 \over q_x} - {y^2 \over q_{y}}  \right) \text{}
\label{epsl}
\end{align}
with $\gamma = \bar{g} - g_2 \, (\bar{k}_{\! \perp} - \bar{k}_{\! \perp s} )^2$ and $\varkappa = \sigma D_u - (\alpha /2) (\bar{k}_{\! \perp}^2 - \bar{k}_{\! \perp s}^2 )$; further, $q_x = \mathrm{w}_x^2 + 
2(g_2 + i \alpha) l$, $q_y = \mathrm{w}_y^2 + 2i \alpha l$ are related to the $1/e^2$-beam widths via $\mathrm{w}^2_{x,y}(l) = |q_{x,y}|^2 / \mathrm{Re}(q_{x,y})$, and the complex 
shift of beam center is given by $ x_c = x_{cr} + i x_{ci} = \alpha l \bar{k}_{\! \perp s} + i g_2 l \, \left( \bar{k}_{\! \perp} - \bar{k}_{\! \perp s} \right) $. We use the following notation; subscript, $s$, 
denotes ($\Omega_s$); otherwise the argument is ($\Omega$).  

From Eq. (\ref{epsl}) the intensity spectrum follows as
\begin{align}
| \tilde{v}_x(x,y,l,\Omega) |^2 & = \frac{ (E_s \tau \mathrm{w}_x \mathrm{w}_y)^2 } { 2 | q_x q_y | } |f(\Omega)|^2  \exp \left( \Gamma l \right) \nonumber \\ 
~& \times \exp \left( - { 2 (x-\xi_{cr})^2 \over \mathrm{w}_x^2(l)} - {2 y^2 \over \mathrm{w}_{y}^2(l)}  \right) \text{.}
\label{epslsq}
\end{align}
Due to contributions from the imaginary parts in the exponent of (\ref{epsl}) the shift of the beam center changes to $\xi_{cr} = x_{cr} + x_{ci} (\mathrm{Im}(q_{x}) / \mathrm{Re}(q_{x}) )$; the gain changes to $\Gamma = 
\bar{g} - g_2 \, (\bar{k}_{\! \perp} - \bar{k}_{\! \perp s})^2 (\mathrm{w}_x^2 / \mathrm{Re}(q_{x}))$. 
Taylor expansion of the gain about $\Omega_s$ yields $\Gamma(\Omega) = \Gamma_s + \Gamma^{\, \prime}_{s} (\Omega-\Omega_s) + (1/2) \, \Gamma^{\, \prime \prime}_{s} (\Omega-\Omega_s)^2$. As a result, the amplified spectrum 
remains Gaussian. Integration over $\Omega$ by using the method of stationary phase results in a spectral $1/e^2$-width $\Delta_{\omega}(l) = 2 / \tau_g(l) $. Here, $\tau_g(l) = (\tau^2 - 
\Gamma^{\, \prime \prime}_{s} l )^{1/2}$ is the gain modified temporal $1/e^2$-duration which corresponds to the actual pulse duration $\tau(l)$ when dispersive effects are small. Finally, integration over 
transverse coordinates yields the amplified seed pulse energy 
\begin{align}
\frac{W_s(l)}{W_s(0)} = \frac{\mathrm{w}_x}{\sqrt{ \mathrm{Re}[q_{x}(\Omega_s)] }} \frac{\tau }{\tau_g(l)} \exp \left( \Gamma_s l + 
\frac{( \Gamma^{\, \prime}_{s} l)^2}{2 \tau_g^2(l) } \right) \text{,}
\label{seeden}
\end{align}
where $W_s(0) = (\pi/2)^{3/2} I_s \tau \mathrm{w}_x \mathrm{w}_y $ and $I_s$ are the initial seed pulse energy and intensity. The spatio-temporal profile could also be calculated by Taylor expanding the exponent 
in Eq. (\ref{epsl}) to second order in $\Omega$ followed by an inverse Fourier transform. Due to the onerous complexity this is not done here. Instead spatio-temporal profiles and $\tau(l)$ are determined numerically 
from Eq. (\ref{epsl}). 

\subsection{Results}

\noindent
KIA operates in the limit where the amplified seed intensity is small compared to the pump peak intensity, so that nonlinear terms in Eq. (\ref{vwaveqt}) are negligible. This is fulfilled for $I_s(l) = I_p/10$ \cite{vampa2017}. 
The corresponding amplified seed pulse energy is $W_s(l) = (\pi/2)^{3/2} I_s(l) \tau(l) \mathrm{w}_x(l) \mathrm{w}_y(l) $, from which together with Eq. (\ref{seeden}) the initial pulse energy and intensity are obtained. 

Efficient amplification requires the seed pulse to stay close to the pump pulse center over the whole amplification distance. This requirement sets a lower limit for pump pulse duration and width, and thereby for the 
minimum pump energy. 

There are four factors that cause an increase in pump energy requirements: i) the inclination between pump and seed pulse axes, resulting in a walk-off $\xi_{cr}$ between beam centers; ii) widening of the seed beam 
widths $\mathrm{w}_{x,y}(l)$ due to diffraction and transverse spectral gain narrowing; iii) a temporal walk-off, $\Delta \beta_1 l$, caused by the difference $\Delta \beta_1 = \beta_{1s} - \beta_{1}$ between seed group 
velocity $\beta_{1s} = [dK_u/ d \Omega](\Omega_s)$ and pump group velocity $\beta_{1}$ defined below Eq. (\ref{dispersion}); iv) lengthening of the seed pulse duration $\tau(l)$ due to spectral gain narrowing and 
dispersive effects. 

These 4 conditions determine the required pump pulse parameters as $\mathrm{w}_{p} = r (\mathrm{w}_x(l) + 0.5 \xi_{cr})$ and $\tau_p = r (\tau(l) + 0.5 |\Delta \beta_1| l ) $; the factor $1/2$ comes from the 
assumption that pump and seed beam centers are aligned at half of the material length; we chose the factor by which the pump beam is wider than the final shifted seed beam as $r=3$ \cite{vampa2017}. 
As a result, the minimum pump energy for KIA to operate efficiently is $W_p = (\pi/2)^{3/2} I_p \tau_p \mathrm{w}^2_{p}$ assuming a radially symmetric transverse pump beam.   

Furthermore, the pump beam radius underlies another restriction; it needs to be wide enough to avoid self focusing. We determine $\mathrm{w}_p$ from the requirement that the material length $ l = l_{sf} / 5$,  where 
$l_{sf} = \mathrm{w}_p (n_p / (2 n_n) )^{1/2}$ is the distance for critical self focusing \cite{boyd2013}. The initial seed beam width $\mathrm{w}_x$ is determined from a solution of 
\begin{align}
l_{sf} \sqrt{2 n_n \over n_p} = \mathrm{w}_{p} = r (\mathrm{w}_x(l) + 0.5 \xi_{cr}) 
\label{wxdet}
\end{align}
with $\mathrm{w}_x(l)$ defined below Eq. (\ref{epsl}); further, we assume $\mathrm{w}_y = \mathrm{w}_x$. Additional parameters to be considered are the nonlinear length $l_n = 2 k_p / k_n^2 = 2 n_p c / 
(n_n \omega_p)$ and dispersive length $l_d = 2 \tau_p^2 / \beta_2$ of the pump pulse. In the limit of strong KIA the nonlinear length is shorter than the material length. As a result, $l_d \gg l$ to avoid pump pulse 
stretching through the combined action of nonlinear phase modulation and dispersion. 

Finally, to keep amplification lengths short it is desirable to use high pump intensities. The obvious limit is the material damage threshold intensity $I_{th} = (2/\pi)^{1/2} F_{th} / \tau_p$
with $F_{th}$ the damage threshold fluence.

\begin{figure}[t]
\hspace*{-0.27cm}\includegraphics[scale=0.45]{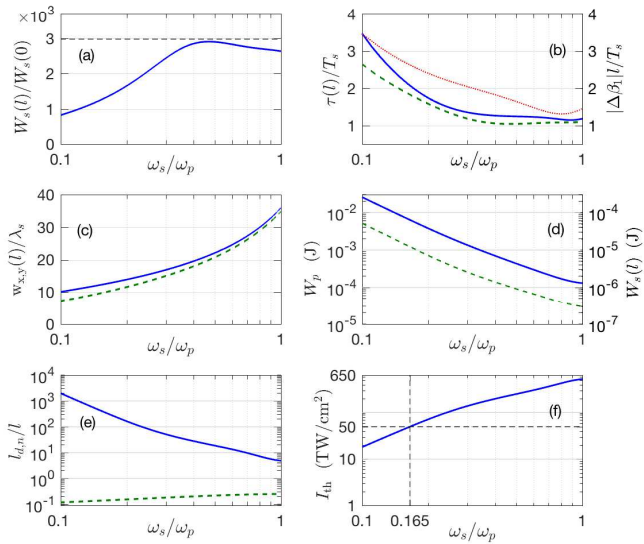}
\caption{KIA of single cycle pulse $\tau(0) = T_s = 2\pi / \omega_s$ in CaF$_2$; $n_2 = 2 \times 10^{-16}$ cm$^2$/W; $I_p = 50$ TW/cm$^2$, $\lambda_p = 0.85\,\mu$m, $l=8/\bar{g}$; $\mathrm{w}_p$ and $\tau_p$, see text 
above Eq. (\ref{wxdet}); $\mathrm{w}_x(0) = \mathrm{w}_y(0)$ is determined from Eq. (\ref{wxdet}). (a) Seed pulse energy increase $W_s(l)/W_s(0)$ from Eq. (\ref{seeden}) versus $\omega_s / \omega_p$; black dashed line 
corresponds to the cw limit $\exp(\bar{g}l) = \exp(8) \approx 3000$. (b) $\tau(l) 
/ T_s$ (blue full), $\tau_g(l)/ T_s$ (green, dashed), and group velocity walk off between pump and seed, $|\Delta \beta_{1}| l/T_s$, versus $\omega_s / \omega_p$ (red, dotted). (c) $w_x(l) / \lambda_s$ (blue, full) 
and $w_y(l) / \lambda_s$ (green, dashed) versus $\omega_s / \omega_p$; initial beam radius is not plotted as $w_y(l) \approx w_x(0) = w_y(0)$. (d) Minimum required pump energy $W_p$ (blue, full), see text 
above Eq. (\ref{wxdet}), and corresponding seed energy $W_s(l)$ (green, dashed) versus $\omega_s / \omega_p$. (e) dispersive length $l_d / l$ (blue, full) and nonlinear length $l_n / l$ (green, dashed) versus 
$\omega_s / \omega_p$. (f) Damage threshold intensity $I_{th}$ versus $\omega_s / \omega_p$; dashed lines indicate where $I_p = I_{th}$. \label{fig3} } 
\end{figure}

\begin{figure}[t]
\includegraphics[scale=0.45]{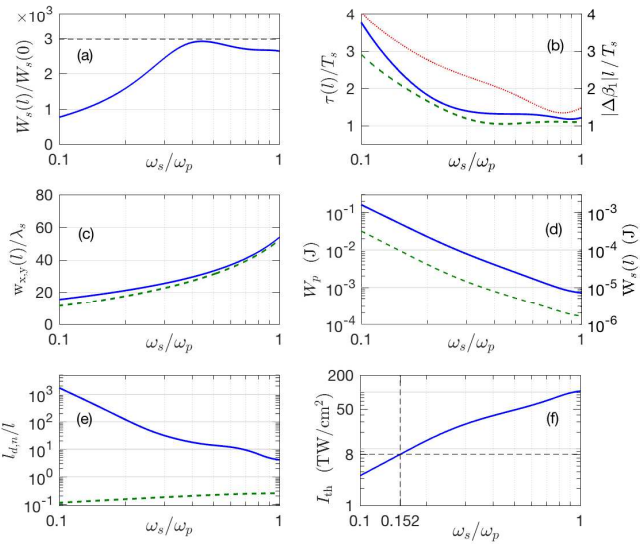}
\caption{KIA of single cycle pulse $\tau(0) = T_s$ in KBr; $n_2 = 6 \times 10^{-16}$ cm$^2$/W; $I_p = 8$ TW/cm$^2$, $\lambda_p = 2.1\,\mu$m, $l=8/\bar{g}$; $\mathrm{w}_p$ and $\tau_p$, see text above Eq. (\ref{wxdet}); 
$\mathrm{w}_x(0) = \mathrm{w}_y(0)$ is determined from Eq. (\ref{wxdet}). (a) Seed pulse energy increase $W_s(l)/W_s(0)$ from Eq. (\ref{seeden}) versus $\omega_s / \omega_p$; black dashed line corresponds to the cw limit 
$\exp(\bar{g}l) \approx 3000$. (b) $\tau(l) / T_s$ (blue full), 
$\tau_g(l)/ T_s$ (green, dashed), and group velocity walk off between pump and seed, $|\Delta \beta_{1}| l/T_s$, versus $\omega_s / \omega_p$ (red, dotted). (c) $w_x(l) / \lambda_s$ (blue, full) and $w_y(l) / 
\lambda_s$ (green, dashed) versus $\omega_s / \omega_p$; initial beam radius is not plotted as $w_y(l) \approx w_x(0) = w_y(0)$. (d) Minimum required pump energy $W_p$ (blue, full), see text above Eq. (\ref{wxdet}), 
and corresponding seed energy $W_s(l)$ (green, dashed) versus $\omega_s / \omega_p$. (e) dispersive length $l_d / l$ (blue, full) and nonlinear length $l_n / l$ (green, dashed) versus $\omega_s / \omega_p$. 
(f) Damage threshold intensity $I_{th}$ versus $\omega_s / \omega_p$; dashed lines indicate where $I_p = I_{th}$. \label{fig4} }
\end{figure}

The quantitative results for finite pulse KIA in CaF$_2$ and KBr are shown in Figs. \ref{fig3} and \ref{fig4}, respectively. We assume a material length $ l = 8 / \bar{g}(\Omega_s)$ corresponding to a plane wave 
amplification factor of $\exp(8) \approx 3000$; the amplifier length $l$ is changed with frequency $\omega_s$ to make the plane wave amplification factor constant for all frequencies. The pump peak intensities for 
CaF$_2$ and KBr are chosen $I_p = 50$ TW/cm$^2$ and $I_p = 8$ TW/cm$^2$, respectively. Following the results of the plane wave analysis above, we chose $\lambda_p = 0.85\,\mu$m and $\lambda_p = 2.1\,\mu$m for 
CaF$_2$ and KBr, respectively. The damage threshold fluence of CaF$_2$ and KBr in the sub-ps pulse duration regime is $F_{th} = 6.7$ J/cm$^2$ and $F_{th} = 3.3$ J/cm$^2$ \cite{gallais2014}, respectively; for $n_2$ 
see caption. We assume single cycle initial seed pulses with $I_s(l) = I_p/10$; initial seed pulse radius is determined from a solution of Eq. (\ref{wxdet}). 

Seed pulse energy increase $W_s(l)/W_s(0)$ in Fig. \ref{fig3}(a) is close to the plane wave value $\exp(\bar{g}l) \approx 3000$ (black, dashed line) for $\omega_s/\omega_p \ge 0.5$ and drops from there; at 
$\omega_s/\omega_p = 0.2$ amplification is still more than a factor of 1000. 

In Fig. \ref{fig3}(b) the $1/e^2$-pulse duration $\tau(l)$ (blue, full) is obtained from transverse space integration over the spatio-temporal intensity profile; the intensity profile is calculated as the absolute 
square of the Fourier transform of Eq. (\ref{epsl}). The pulse duration $\tau(l)$ is compared to $\tau_g(l)$ (green, dashed) which is the gain widened pulse duration defined below Eq. (\ref{epslsq}); it is obtained 
from the spectral width and does not contain dispersive widening. Comparison shows that up to $\omega_s/\omega_p \approx 0.3$ amplification of single cycle pulses is possible and that the influence of dispersive 
effects is weak; even at $\omega_s/\omega_p = 0.2$ amplification of two cycle pulses is still feasible. Below that the pulse duration rises quickly due to a mixture of gain and dispersive widening. 
Finally, the red dotted line indicates the shift between peak of seed and pump pulse due to group velocity mismatch. 

Widening of the seed beam radius is not dramatic, as can be seen in Fig. \ref{fig3}(c). This is due to the fact that a large pump beam radius is required to avoid self-focusing. This results in a large seed beam 
radius, as in our above design considerations the seed radius increases proportional with the pump radius. In general, it is desirable to choose the seed beam radius as large as possible to optimize energy extraction 
from the pump beam. We find that (green, dashed) $\mathrm{w}_y(l) \approx \mathrm{w}_x(0) = \mathrm{w}_y(0)$ which is why the initial pulse radii are not plotted. Amplification moderately widens $\mathrm{w}_x(l)$ 
(blue, full), as defined below Eq. (\ref{epsl}), and results in a beam asymmetry which is weak over most of the frequency range. 

\begin{figure}\vspace*{-0.2cm}
\hspace*{-0.2cm}\vspace*{-0.2cm}\includegraphics[height=5.5cm, width=9.6cm]{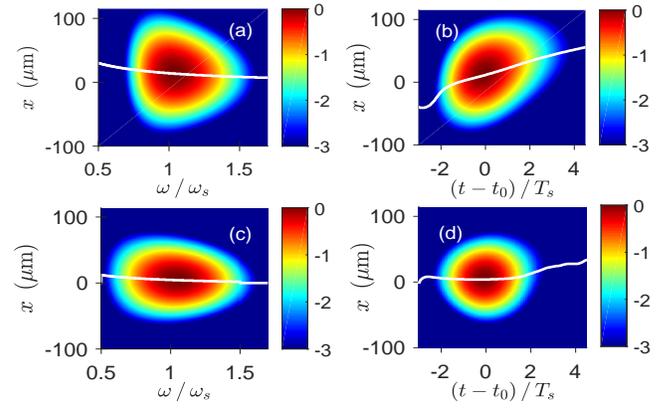}
\caption{ Spatio-spectral (a,c) and spatio-temporal (b,d) intensity profiles of seed pulses amplified in CaF$_2$ for $\omega_s / \omega_p = 0.2, 0.4$, respectively; parameters are the same as in Fig. 
\ref{fig3}; peaks are normalized to unity; time is given with reference to time $t_0$ of the pulse peak and normalized to the optical cycle $T_s$. The white lines indicate the transverse pulse maximum.\label{fig5} }
\end{figure}

In Fig. \ref{fig3}(d) the minimum pump pulse energy needed for KIA to work and the corresponding amplified seed pulse energy are plotted versus $\omega_s / \omega_p$. Naturally, higher seed energies can be 
obtained when more pump energy is available. At $\omega_s / \omega_p = 0.2$ we find $W_p = 4$ mJ which is comfortably available in Ti:sapphire laser systems. The pump energy is larger than the final seed energy by a 
factor of about $400-500$.\\
The nonlinear length $l_n$ (green dashed) is shorter than the amplifier length, see Fig. \ref{fig3}(e). The dispersive length $l_d$ (blue, full) is between two to four orders of magnitude longer than the medium length so 
that no significant pump pulse distortions are expected through the interplay of Kerr nonlinearity and group velocity dispersion.\\
Finally, Fig. \ref{fig3}(f) shows the damage threshold intensity $I_{th}$ for a pump pulse with pulse duration $\tau(l)$. The dashed line indicates the value of $\omega_s / \omega_p = 0.165$ at which $I_p = 
I_{th}$. As a result, we can conclude that amplification for a wavelength range between $\lambda_s = 0.85\,\mu$m and $\lambda_s \approx 5.2\,\mu$m is possible. The damage intensity presents a main 
limitation in extending KIA to even longer wavelengths. Reducing $I_p$ does not help. This results in an increase of material length $l$ to achieve the same amplification; longer $l$ results in larger $\tau(l)$, 
and in an enhanced walk-off, which results in turn in longer pump pulse duration $\tau_p$ and reduced damage threshold intensity.\\
Figure \ref{fig4} shows the results for KBr. The results are qualitatively similar to what was found for CaF$_2$ in Fig. \ref{fig3}; therefore we focus on a discussion of Fig. \ref{fig4}(d) and (f). The minimum 
required pump energy $W_p \approx 20$ mJ at $\omega_s / \omega_p = 0.2$. This is in the range of what can be achieved by current state of the art Ho:YAG femtosecond amplifier systems operating at wavelengths 
$\lambda_p = 1.9 - 2.1\,\mu$m \cite{malevich2013}. The corresponding seed amplified energy is $W_s \approx 50\,\mu$J. From \ref{fig4}(f) we find that KIA is possible for $\omega_s / \omega_p > 0.152$ corresponding 
to a maximum seed wavelength of $\lambda_s \approx 14\,\mu$m.\\
Finally, it is interesting to look at the quality of the amplified pulses. Again our two systems behave fairly similar, which is why we show only the results for CaF$_2$ and $\lambda_p = 0.85\,\mu$m; for other 
parameters see Fig. \ref{fig3}. In Fig. \ref{fig5}(a),(c) the spatio-spectral intensity profile $|\tilde{\varepsilon}_x(x,y=0,z=l,\omega)|^2$ is plotted for $\omega_s/\omega_p = 0.2, 0.4$, respectively; 
Figures \ref{fig5}(b),(d) show the corresponding spatio-temporal profiles $|\varepsilon_x(x,y=0,z=l,t)|^2$; peak values are normalized to unity. The spectrum peak is shifted off $\omega / \omega_s = 1$ towards 
higher frequencies. This comes from the fact that the blue part of the seed spectrum is amplified more strongly, as for $\omega_s/\omega_p \le 0.4$ $\bar{g}$ increases towards higher seed frequencies. Further, 
the spectrum exhibits some asymmetry which is not contained in the quadratic expansion of $\Gamma$ below Eq. (\ref{epslsq}); accounting for it analytically would require expansion to third order. 
The fact that maximum gain $\bar{g}$ is experienced at finite transverse wavevector $\bar{k}_{\perp}(\Omega)$, the value of which is frequency dependent, results in a Gaussian pulse in space domain shifted by 
$x_c$, see Eq. (\ref{epsl}). The real part of the shift $\xi_{cr}$ manifests as an off-axis shift of the pulse center, see the white line in \ref{fig5}(a),(c); the shift changes slightly with frequency as a result 
of the angular chirp, i.e. each frequency experiences optimum amplification at a slightly different angle. The angular chirp needs to be compensated, as otherwise the frequency dependent shift of the pulse center 
will continue growing during free space propagation \cite{vampa2017}, resulting in a degradation of the pulse quality. The imaginary part $x_{ci}$ has an effect on the spatio-temporal pulse in \ref{fig5}(b),(d). 
It creates an x-dependent group velocity component which skews the pulse in the $x-t$ plane. The pulse distortion becomes pronounced for $\omega_s / \omega_p \le 0.2$ and is negligible for $\omega_s / \omega_p 
\ge 0.35$. 

\section{Conclusion}

\noindent
We have introduced a new concept for amplification of mid infrared pulses based on the Kerr instability. Our proof-of-principle theoretical analysis of KIA in CaF$_2$ and KBr crystals demonstrates the potential to 
amplify pulses in the wavelength range $\approx 1-14\,\mu$m. Whereas plane wave amplification in KBr extends to $40\mu$m, material damage limits finite pulse KIA to about $14\mu$m. There, seed pulse output energies 
in the $50\,\mu$J range appear feasible with a ratio of pump to seed pulse energy in the range 400-500. Our numbers are comparable to the performance of optical parametric amplifiers.\\
The biggest three advantages of KIA are the capacity for single cycle pulse amplification, that it is intrinsically phase matched, and its simplicity and versatility; Kerr materials are more easily 
available than infrared materials with second order nonlinearity. Further, amplifier wavelength can be selected by simply changing the angle between pump and seed beam. The biggest drawback is an angular chirp 
acquired during amplification that needs to be controlled. There exist methods to that end, from a simple prism to more sophisticated techniques \cite{mendoza2010}. Alternatively, it should also be possible to 
identify favorable materials that minimize the angular chirp, as the angular chirp is greatly influenced by the frequency dependence of the refractive index.\\
The results shown here are promising, but most likely still far from optimum. There is a huge parameter space to be explored, such as all potential infrared crystals. Further, KIA can be optimized by determining 
favorable optical properties (e.g. refractive index) from our theory and then designing corresponding (meta) materials. Moreover, restrictions of the amplification range arising from material damage can be 
mitigated by crystal cooling and parameter optimization. Finally, the KIA profile is of Bessel-Gaussian nature. Therefore KIA should lend itself naturally to the amplification of Bessel-Gauss beams. \\See Supplement 1 for
supporting content.

\clearpage
\section{Supplementary Material}
\noindent

\subsection{Limiting cases of KIA theory}
\label{sec1}

\noindent
In the limits of $\mathbf{k}_{\perp} = 0$ and $\Omega=0$, Eq. (\ref{KOmg}) reduces to the temporal modulation instability \cite{agrawal2012}, and the spatial filamentation instability 
\cite{bespalov1966}, respectively. For $\Omega/\omega_p \ll 1$ and $n_n \ll 1$ we can approximate $D_u \approx \beta_1 \Omega$, $D_{g} \approx \beta_2 \Omega^2 / 2$, $\sigma \approx 1$, 
$k_p^2 - (\sigma D_u)^2 \approx k_p^2$, $\delta_{\! \perp}^2 \approx k_n^2(\omega_p)$. By using the approximation below Eq. (\ref{quartic}) we find $ \sigma^2-1 \approx (k_n(\omega_p)/k_p)^2 + 
2 D_g/k_p$ and obtain 
\begin{align}
g \approx \sqrt{ \left( {k_n^2(\omega_p) \over k_p} \right)^2 - \left( {k_n^2(\omega_p) \over k_p} + {\beta_2 \Omega^2} - {k_{\perp}^2 \over k_p}\right)^2 } \text{.}
\label{gainapprox}
\end{align}
By setting $\Omega = 0$ in Eq. (\ref{gainapprox}), a relation for the filamentation instability is obtained in agreement with \cite{bespalov1966}. By setting $k_{\perp} = 0$ and by introducing the fiber nonlinear 
coefficient $\gamma = n_2 \omega_p / (2n_p c A_{\rm eff})$ we can express the nonlinear term as $k_{n}^2(\omega_p) / k_p = 2 \gamma P_p$. Here $I_p = P_p / A_{\rm eff}$, $A_{\rm eff}$ is the effective fiber 
pulse area, and $P_p$ the pump peak power. The equation resulting from Eq. (\ref{gainapprox}) agrees with the gain for modulation instability in fibers \cite{agrawal2012},   
\begin{align}
g \approx \sqrt{ (\beta_2 \Omega^2)^2 + 4\gamma P_p \beta_2 \Omega^2} \text{.}
\label{gainmodi}
\end{align}
Finally, note that we have defined the total refractive index, $(n_p^2 + n_n)^{1/2}$, differently to Ref. \cite{agrawal2012}, where $n_p + n_n$ is used; as a result $2 \gamma n_p$ corresponds to $\gamma$ defined in 
Ref. \cite{agrawal2012}. 

\subsection{Summary of definitions and parameters}
\label{sec2}
\noindent
In the following a summary of the definintions and variables used in this work is given. For variables defined in the text we give the equation number and use $\uparrow$ or $\downarrow$ to indicate the location of the defnition with regard to the equation number.\\
\begin{table}[!htbp]
\begin{tabular}[t]{lcl}
Location					&Variable								&$\,\,\,\,\,\,\,\,\,\,\,\,$Description\\
\hline
Section \rm{II}\vspace{1mm}\\				
$\ua$(\ref{vwaveqt})		&$\mathbf{E}(\mathbf{x},t)$				&$\mathbf{E}(\mathbf{x},t)=\boldsymbol{\varepsilon}(\mathbf{x},t)+$\\ 
						&									&$\hat{\mathbf{x}} E_{p}\exp(i\omega_{\rm p} t-i k_{p} z) +  {\rm c.c.}$\\
$\ua$(\ref{vwaveqt})		&${E_{p}}$							&Pump electric field amplitude\\
$\ua$(\ref{vwaveqt})		&$\omega_{p}$						&Pump angular frequency\\
$\da$(\ref{vwaveqt})		&$k_{p}$								&Pump wavevector\\
$\ua$(\ref{vwaveqt})		&$\bs{\varepsilon}(\mathbf{x},t)$ 			&Small perturbation (seed)\\	
$\da$(\ref{vwaveqt})		&$n(\omega)$							&Linear refractive index\\
$\da$(\ref{vwaveqt})		&$n_{2}$								&Optical Kerr nonlinear index\\	
$\da$(\ref{vwaveqt})		&$n_{n}$								&$n_{n} = n_2I_{p}$\\	
$\da$(\ref{vwaveqt})		&$I_{p}$								&Pump intensity\\
$\ua$(\ref{vwaveq})		        &$\bs{v}(\mathbf{x},t)$					&$\bs{v}(\mathbf{x},t)=\bs{\varepsilon}(\mathbf{x},t)\exp(-i\omega_{\rm p}t + ik_{\rm p}z)$\\		
$\ua$(\ref{vwaveq})			&$\bs{k}_{\perp}$					        &Transverse wavevector\\
$\ua$(\ref{vwaveq})			&$\tilde{\bs{v}}(z,\bs{k}_\perp,\Omega)$ 	&Fourier transform of $\bs{v}(\mathbf{x},t)$\\
$\da$(\ref{vwaveq})			&$k_v$								&$k_v=\sqrt{k^2 + 2k_n}$\vspace{1.7mm}\\
\hline												
\end{tabular}
\end{table}
\begin{table}[!htbp]
\begin{tabular}[t]{lcl}
Location					&Variable							        &$\,\,\,\,\,\,\,\,\,\,\,\,$Description\\
\hline
Section \rm{II}\vspace{3.5mm}\\
$\da$(\ref{vwaveq})			&$k$								&$k=n(\omega)\omega/c$\\
$\da$(\ref{vwaveq})			&$k_{n}$								&$k_n=\sqrt{n_n}\omega/c$\\
$\da$(\ref{vwaveq})			&$k_\perp^2$							&$k_\perp^2=k_x^2 + k_y^2$\\
$\da$(\ref{vwaveq})			&$\tilde{\bs{v}}_{(-)}^*$ 					&$\tilde{\bs{v}}_{(-)}^*=\tilde{\bs{v}}^*(-\Omega)$\\
$\da$(\ref{vwaveq-})	 	&$\eta$								&$\eta=\sqrt{n^2+2n_n}$\\	
$\da$(\ref{vwaveq-})		&$\eta_{p}$							&$\eta_{p}=\eta(\omega_{p})=\eta(\Omega=0)$\\
$\ua$(\ref{dispersion})		&$\Delta\eta(\Omega)$					&$\Delta\eta(\Omega) = \eta(\omega_{p}+\Omega) - \eta_{p}$\\
$\ua$(\ref{dispersion})		&$\eta_{g}(\Omega)$					&$\eta_{g}(\Omega)=[\Delta\eta(\Omega)+\Delta\eta(-\Omega)]/2$\\
$\ua$(\ref{dispersion})		&$\eta_{u}(\Omega)$					&$\eta_{u}(\Omega)=[\Delta\eta(\Omega)-\Delta\eta(-\Omega)]/2$\\
(\ref{Dg})    				&$D_{g}(\Omega)$						&Even dispersion function\\ 
(\ref{Du})				        &$D_{u}(\Omega)$					        &Odd dispersion function\\
$\da$(\ref{dispersion})		&$\beta_1$							&$\beta_1 = [dk/d\omega](\omega_p)$\\
$\da$(\ref{dispersion})		&$\beta_2$							&$\beta_2 = [d^2k/d\omega^2](\omega_p)$\\
$\da$(\ref{quartic})			&$\sigma$							&$\sigma = [k_{v}(\omega_{p}) + D_{g}]/k_{p}$\\
$\da$(\ref{quartic})			&$\sigma^2-1$						&$\sigma^2-1\approx (k_n(\omega_p) / k_p)^2 + 2 D_{\rm g} / k_{\rm p}$\\
$\ua$(\ref{KOm})			&$\kappa_\perp$						&$\kappa_\perp=\sqrt{(k_p^2-D_u^2)(\sigma^2-1)}$\\		
$\ua$(\ref{KOm})			&$K_v$								&$K_v = K_u + K_g$\\
(\ref{KOmu})				&$K_u$								&$K_{u} = -\sigma D_{u}\left[1-\frac{(\kappa_\perp^2 - k_\perp)^2}{2(k_{p}^2 - \sigma^2D_{ u}^2)}\right]$\\
(\ref{KOmg})				&$K_{g}$								&$K_{g} = -\frac{k_{p}\sqrt{(\kappa_\perp^2 - k_\perp)^2 - \delta_\perp^4}}{2(k_{p}^2 - \sigma^2D_{u}^2)}$\\
$\ua$(\ref{kpmax})			&$g$								&Intensity gain, $g=-2\rm{Im}(K_g)$\\	
(\ref{kpmax})				&$\bar{k}_\perp$						&Transverse wavevector for max gain\\
(\ref{gmax})				&$\bar{g}$							&Max intensity gain\\
(\ref{kwidth})				&$\delta_\perp$						&Transverse instability\\ 
						&									&half-width\vspace{2.5mm}\\
\hline 		
Section \rm{III}\vspace{2.5mm}\\			
$\da$(\ref{pwfourier})		&$l$									&Kerr material length\\
$\da$(\ref{pwfourier})		&$\bar{k}_{\perp s}$					&$\bar{k}_{\perp s} = \bar{k}_{\perp}(\Omega_s)$\\ 
$\da$(\ref{pw})			&$\mathbf{K}(\Omega_s)$				&Instability wavevector \\
$\da$(\ref{pw})			&$K_z$ 								&$K_z = k_p + \sigma D_u$\\
$\da$(\ref{pw})			&$K_{zs}$								&$K_{zs} = K_z(\Omega_s)$\\
$\da$(\ref{pw})			&$E_s$ 								&Seed electric field strength\\
$\da$(\ref{pw})			&$\omega_s$ 							&$\omega_s = \omega_p + \Omega_s$\\
$\da$(\ref{pw})			&$\nu_s$ 							&Seed frquency, $\nu_s=\omega_s/(2\pi)$\\
(\ref{thetas})				&$\theta_s$ 							&$\theta_s=\arctan\left(\bar{k}_{\perp s}/K_{zs}\right)$\\
$\da$(\ref{thetas})			&$\lambda_{p,s}$ 						&Pump, seed wavelength\vspace{6.9mm}\\
\hline
\end{tabular}
\end{table}
\begin{table}[!htbp]
\begin{tabular}[t]{lcl}			
Location					&Variable								&$\,\,\,\,\,\,\,\,\,\,\,\,$Description\vspace{0.2mm}\\
\hline
Section \rm{IV}\vspace{2mm}\\
$\ua$(\ref{vseed0})			&$\mathrm{w}_{x,y}(0)$					&$1/e^2$ initial seed widths,\\
						&									&$\mathrm{w}_{x,y}(0) = \mathrm{w}_{x,y}$\\	
$\ua$(\ref{vseed0})			&$\Delta_{x,y}$						&$\Delta_{x,y}=2/\mathrm{w}_{x,y}$\\
$\ua$(\ref{vseed0})			&$\tau(0)$							&$1/e^2$ initial seed duration,\\ 
						&									&$\tau(0)=\tau=T_s$\\
$\ua$(\ref{vseed0})			&$\Delta_{\omega}$					&$\Delta_{\omega}=2/\tau$\\
(\ref{vseed0})				&$\tilde{v}_{x}(0)$						&Initial Fourier-transformed\\ 
						&									&Gaussian seed pulse\\
$\da$(\ref{vseed0})			&$f(\Omega)$							&$f(\Omega)=\exp\left(-(\Omega-\Omega_s)^2/\Delta_\omega^2\right)$	\\
(\ref{intamp})				&$g_2$								&$g_2 = \frac{2k_{p}\bar{k}_\perp^2}{\delta_\perp^2(k_{p}^2 - \sigma^2D_{u}^2)}$\\
(\ref{epslft})				&$\tilde{v}_x(\bs{k}_\perp,l,\Omega)$		&Fourier beam amplitude at $l$ \\
$\da$(\ref{epslft})			&$\alpha$							&$\alpha = \frac{\sigma D_{u}}{k_{p}^2 - \sigma^2D_{u}^2}$\\	
(\ref{epsl})				&$\tilde{v}_x(x,y,l,\Omega)$				&Amplified, shifted seed\\
$\da$(\ref{epsl})			&$\gamma$							&$\gamma = \bar{g} - g_2(\bar{k}_\perp - \bar{k}_{\perp s})^2$\\	
$\da$(\ref{epsl})			&$\varkappa$							&$\varkappa = \sigma D_u - \alpha/2(\bar{k}_\perp^2 - \bar{k}_{\perp s}^2)$ \\
$\da$(\ref{epsl})			&$q_x$								&$q_x = \mathrm{w}_x^2 + 2(g_2+i\alpha)l$ \\
$\da$(\ref{epsl})			&$q_y$								&$q_y = \mathrm{w}_y^2 + 2i\alpha l$ \\
$\da$(\ref{epsl})			&$\mathrm{w}_{x,y}(l)$					&$\mathrm{w}_{x,y}(l) = \frac{\left|q_{x,y}\right|}		{\sqrt{\mathrm{Re}\left(q_{x,y}\right)}}$ \\
$\da$(\ref{epsl})			&$x_c$								&Complex seed center,\\
						&									&$x_c = x_{cr}+ix_{ci}$\\	
$\da$(\ref{epsl})			&$x_{cr}$								&$x_{cr} = \alpha l\bar{k}_{\perp s}$\\
$\da$(\ref{epsl})			&$x_{ci}$								&$x_{ci} = g_2l(\bar{k}_\perp-\bar{k}_{\perp s})$\\	
(\ref{epslsq})				&$|\tilde{v}_x(l)|^2$					&Intensity spectrum of amplified\\																		        				&									&complex-shifted Gaussian seed\vspace{0.635mm}\\			
$\da$(\ref{epslsq})			&$\xi_{cr}$							&$\xi_{cr} = x_{cr} + x_{ci}(\rm{Im}(q_x)/\rm{Re}(q_x))$\\	
$\da$(\ref{epslsq})			&$\Gamma$							&$\Gamma=\bar{g} - g_2(\bar{k}_\perp-\bar{k}_{\perp s})^2\frac{\mathrm{w}_x^2}{\mathrm{Re}			(q_x)}$\\
$\da$(\ref{epslsq})			&${\Gamma_s}$						&${\Gamma_s}={\Gamma}(\Omega_s)$\\
$\da$(\ref{epslsq})			&$\tau_g(l)$							&$\tau_g(l) = \sqrt{\tau^2 - \Gamma_{s}''l}$\\
$\da$(\ref{epslsq})			&$\Delta_\omega(l)$					&$\Delta_\omega(l)=2/\tau_g(l)$\\	
$\ua$(\ref{seeden})			&$\tau(l)$							&Actual pulse duration after $l$\\	
(\ref{seeden})				&$W_{s}(l)$							&Amplified seed pulse energy\\
$\da$(\ref{seeden})			&$W_{s}(0)$							&Initial seed pulse energy\\	
						&									&$W_s(0) = (\pi/2)^{3/2} I_s \tau \mathrm{w}_x \mathrm{w}_y$\\	
$\da$(\ref{seeden})			&$I_s$								&Initial seed intensity\\					
$\ua$(\ref{wxdet})			&$\Delta\beta_1$						&Group velocity mismatch,\\	
						&									&$\Delta\beta_1=\beta_{1s}-\beta_1$\\						
$\ua$(\ref{wxdet})			&w$_p$								&Pump width, \\
						&									&$\mathrm{w}_p=r\left(\mathrm{w}_x(l)+0.5\xi_{cr}\right)$\vspace{2mm}\\
\hline
\end{tabular}
\end{table}
\begin{table}[!htbp]
\begin{tabular}[t]{lcl}			
Location					&Variable								&$\,\,\,\,\,\,\,\,\,\,\,\,$Description\vspace{0.2mm}\\
\hline
Section \rm{IV}\vspace{2mm}\\
$\ua$(\ref{wxdet})			&$\tau_p$							&Pump duration, \\
						&									&$\tau_p=r\left(\tau(l)+0.5|\Delta\beta_1|l\right)$\\
$\ua$(\ref{wxdet})			&$r$									&Pump to seed width ratio, $r=3$\\
$\ua$(\ref{wxdet})			&$W_p$								&Pump energy,\\
						&									&$W_p = (\pi/2)^{3/2} I_p \tau_p \mathrm{w}^2_{p}$\\
$\ua$(\ref{wxdet})			&$l_{sf}$								&Self-focusing length, \\	
						&									&$l_{sf}=\mathrm{w}_p\sqrt{n_p/(2n_n)}$\\
$\da$(\ref{wxdet})			&$l_{n}$								&Nonlinear length, $l_{n}=2n_pc/(n_n\omega_p)$\\
$\da$(\ref{wxdet})			&$l_{d}$								&Dispersion length, $l_{d}=2\tau_p^2/\beta_2$\\
$\da$(\ref{wxdet})			&$F_{th}$								&Damage threshold fluence\\
$\da$(\ref{wxdet})			&$I_{th}$								&Damage threshold intensity,\\
						&									&$I_{th}=\sqrt{2/\pi}F_{th}/\tau_p$\vspace{3.5mm}\\	
\hline								
\end{tabular}
\caption{Summary of the variables, their locations, and definitions used in this work.}\vspace{5mm}
\end{table}




\end{document}